# Financial knowledge and borrower discouragement

**David Aristei**
University of Perugia

**Manuela Gallo**
University of Perugia

**Raoul Minetti**
Michigan State University

*Preliminary version—please do not cite without authors' permission*

**Abstract**
This study provides first empirical evidence on the impact of entrepreneurs' financial knowledge on borrower discouragement. Using novel survey data on Italian micro-enterprises, we find that less financially knowledgeable entrepreneurs are more likely to be discouraged from applying for new financing, due to higher application costs and expected rejection. Our main results are robust to several sensitivity checks, including accounting for potential endogeneity. Furthermore, we show that the observed self-rationing mechanism is rather inefficient, suggesting that financial knowledge might play a key role in reducing credit market imperfections.

# 1. Introduction

Limited access to external financing poses significant challenges to micro, small, and medium enterprises (MSMEs), hindering their investment opportunities and growth potential.

Extant literature has analysed credit rationing mainly from a supply-side perspective, focusing on loan rejection and quantity restrictions in credit provision to firms. Information asymmetries are at the hearth of these issues, as they prevent lenders to distinguish good and bad borrowers, leading to a market equilibrium characterized by adverse selection and credit rationing (Stiglitz and Weiss, 1981; Ferrando et al., 2017). Financing constraints might as well be the result of a self-rationing decision by firms that need credit but are discouraged from applying in expectation of denial (Jappelli, 1990; Levenson and Willard, 2000). Kon and Storey (2003) show theoretically that, under imperfect information, firms are screened with error and incur financial, in-kind, and psychological application costs; thus, potential borrowers may choose not to apply for credit because they feel they will be rejected. Discouragement is a widely spread phenomenon in developed and emerging economies (Cowling and Sclip, 2023) and might be an inefficient and economically costly self-rationing mechanism. Recent studies have in fact provided evidence that a significant part of discouraged firms would have obtained credit if they had applied (Cole and Sokolyk, 2016; Ferrando and Mulier, 2022; Wernli and Dietrich, 2022).

Previous research has drawn attention on the potential role of financial literacy on the demand-side of MSMEs' access to finance (Naegels *et al.*, 2022; Basha et al., 2023). Lack of financial knowledge and skills may in fact increase borrower discouragement by reducing firms' ability to go through loan application procedures and properly self-assess their own creditworthiness (Kon and Storey, 2003; Balana *et al.*, 2022). Only few studies have attempted to empirically assess the effect of financial competencies on firms' access to credit. Xu *et al*. (2020) show that owners' financial literacy strongly enhances businesses' credit accessibility. Nguyen *et al*. (2021) focus on the impact of formal education, used as an imperfect proxy for financial literacy, on borrower discouragement and find that entrepreneurs without a university-level degree are more likely to self-ration due to high application costs.

This study is the first to provide empirical evidence on the impact of entrepreneurs' financial knowledge on borrower discouragement. Exploiting novel survey data on Italian micro-enterprises, we find that less financially knowledgeable entrepreneurs are more likely to be discouraged from applying for credit, as they face higher application costs and because they expect rejection. We further show that the observed self-rationing mechanism is inefficient, suggesting that financial knowledge might significantly contribute to reduce credit market imperfections. Italy is particularly relevant to the aims of our analysis, as it is characterized by a bank-based financial system, where financial intermediaries play a crucial role in providing access to debt financing for MSMEs. Moreover, the share of entrepreneurs with adequate financial skills in Italy is still quite low (D'Ignazio et al., 2022), making the Italian context ideal for analysing the impact of financial knowledge on borrower discouragement.

The paper proceeds as follows. Section 2 describes data and methods. Empirical results are presented and discussed in Section 3. Section 4 provides concluding remarks.

## 2. Data and methods

We use data from the "Financial literacy and digitalization of small businesses in Italy" survey, carried out in 2021 by the Bank of Italy on a representative sample of about 2,000 Italian non-financial firms with less than 10 employees. The survey focuses on entrepreneurs' financial literacy, based on the OECD/INFE harmonized methodology (OECD, 2020), and provides information on firm characteristics, use of financial products, management of business finances, and access to credit.[1]

In line with previous literature (Cole and Sokolyk, 2016), we first identify firms having a demand for finance as those that, since the start of the COVID-19 pandemic, either applied for a loan or did not apply despite they needed new financing (*Need*). We then classify as discouraged firms that needed a loan but did not apply due to too complex application procedures or because they did not think the loan would be approved (*Discouraged*). Table 1 shows that 68.2% of the firms in the sample needed additional financing, while 12.6% of them were discouraged from applying.

Our main explanatory variable is entrepreneurs' objective *Financial knowledge*, measured as the number of correct answers to seven test-based questions on fundamental financial concepts (i.e., simple interest, compound interest, risk-return relationship, inflation, interest on loans, credit ratings, risk diversification), essential for making informed investment and borrowing decisions.[2] From Figure 1, we notice that more than 40% of the entrepreneurs answer correctly at least 6 questions. The Figure also shows that the incidence of borrower discouragement is significantly lower among more financially knowledgeable entrepreneurs.

Our empirical models also include a rich set of firm and entrepreneur characteristics, as well as sector and region fixed effects; Table 1 reports complete variable definitions.

[Table 1 here]

[Figure 1 here]

To analyse discouragement probability, we need to control for the potential selection bias arising from the fact that the subsample of firms needing new financing might be non-random. We thus consider the following probit model with sample selection:

$Need_i = 1(\alpha_1 FK_i + z_i'\alpha_2 + u_i > 0)$ (1)

$Discouraged_i = 1(\beta_1 FK_i + x_i'\beta_2 + \varepsilon_i > 0)$, observed only if $Need_i = 1$ (2)

---

[1] The estimation sample consists of 1821 firms for which we have complete data for all the variables used in the analysis.
[2] The exact wording of the questions is reported in Table S1 in the Supplementary Appendix.

where $FK_i$ denotes financial knowledge and $x_i$ and $z_i$ are vectors of controls. To improve identification, we exclude from $x_i$ the indicators *Low pre-pandemic liquidity*, *Liquidity decrease*, and *Profits decrease*, assuming that they affect only the probability of needing financing. The errors $u_i$ and $\varepsilon_i$ are assumed to follow a standard bivariate normal distribution, with correlation $\rho$. If $\rho \neq 0$, the sub-sample of credit needing firms is non-random and standard probit results will be biased.

**3. Results**

*3.1 Main results*

Table 2 presents results in terms of estimated average marginal effects (AMEs) on the probability of needing new financing and on the conditional probability of being discouraged. From the Table, we first notice that error correlation $\rho$ is highly significant, suggesting that non-random selection must be accounted for to avoid biased results.

The novel result that our analysis adds to the literature on firms' access to finance relates to the key role of financial knowledge in alleviating borrower discouragement. The estimated AME of $FK_i$ on the conditional probability of not applying for a loan is statistically significant at the 1% level and indicates that a one-unit increase in the financial knowledge score reduces discouragement probability by about 1.5 percentage points. Conversely, financial knowledge does not affect the probability of needing additional financing. This evidence provides first empirical support to the theoretical insights of previous research and suggests that lack of adequate financial knowledge represents a significant demand-side barrier for micro-enterprises' access to finance.

[Table 2 here]

With respect to the other control variables, we find that one-person businesses, firms with an annual turnover lower than €100,000, and those located in Central and Southern Italy are significantly more likely to be discouraged. Enterprises that received advice to accessing external finance not only have higher financing needs, but also have a lower discouragement probability, as they benefit from lower application costs due to improvements in their ability to prepare loan applications (Kon and Storey, 2003). Differently from previous research (Bertrand *et al.*, 2022), we also find that female entrepreneurs in Italy have a significantly lower probability of being discouraged than their male counterparts. Finally, we show that entrepreneurs with an upper secondary education are less likely to be discouraged, while higher qualifications do not impact borrower discouragement. This result highlights that financial knowledge exerts a separate significant effect on discouragement, while the effect of formal education is minor.

*3.2 Robustness and additional analyses*

We carry out several analyses to assess the validity of our main findings (Table 3).³

First, we address potential endogeneity concerns and extend the baseline model by adding a reduced form equation for $FK_i$ as a function of $z_i$ and additional exogenous instrumental variables. Following previous literature (Lusardi and Mitchell, 2014), we use a dummy equal to one if the entrepreneur received financial education at school and a dummy indicating whether at least one of the entrepreneur's parents owns or owned a business in the past. The choice of these instruments is based on the assumption that exposure to financial education in school and parental business experience affect entrepreneurs' financial knowledge, but not directly impact on their loan demand behaviour. Results (panel *a*) of Table 3) support our identification strategy and indicate that financial knowledge is exogenous with respect to loan demand behaviour, allowing for a causal interpretation of our main empirical findings. Estimated AMEs are in line with baseline results and lend support to the significant impact of financial knowledge on reducing borrower self-rationing.

[Table 3 here]

Then, we disentangle the impact of financial knowledge on the different reasons of borrower discouragement (panel *b*) of Table 3). Results point out that financial knowledge significantly reduces the probability of being discouraged due to both complex application procedures and anticipated rejection.⁴ Coherently with the predictions of the theoretical model of Kon and Storey (2003), less financially knowledgeable entrepreneurs face higher sunk application costs and are also less confident about a positive outcome of their loan application, becoming more likely to self-ration themselves.

As a final robustness check, we extend the baseline model by including an additional measure of financial knowledge, based on the number of correct answers to four questions on basic concepts related to business finance (i.e., dividends, equity-control relationship) and accounting (i.e., balance sheets, ROA). Results (panel *c*) of Table 3) confirm that knowledge of fundamental financial concepts significantly reduces the conditional probability of not applying for credit, while business finance and accounting knowledge does not affect borrower discouragement. Interestingly, entrepreneurs with higher knowledge of basic business finance and accounting concepts are significantly less likely to need a loan, possibly suggesting that they may have adequate internal financial resources and/or rely more heavily on other sources of external finance.

---

³ Complete results of these additional analyses are presented in Tables S2 to S4 in the Supplementary Appendix.
⁴ Interestingly, this disaggregated analysis allows pointing point that the lower discouragement probability of female entrepreneurs is entirely driven by women's lower likelihood of not applying due to too complex application procedures and not by differences in loan rejection expectations (Table S3 in the Supplementary Appendix).

*3.3. The inefficiency of borrower discouragement*

We finally assess whether discouragement is an inefficient mechanism that may lead creditworthy firms to self-ration and be left out of the credit market. To this aim, we carry out a counterfactual analysis to gauge the probability that the discouraged borrowers in our sample would have been approved had they applied for a loan. Following Ferrando and Mulier (2022), we thus estimate a model for the probability of having a loan application approved, predict the probability of obtaining credit for approved and discouraged firms, and compare the predicted values for the two groups.[5] Figure 2 plots the cumulative distribution function of the estimated approval probability of discouraged borrowers (vertical axis) relative to that of approved applicants (horizontal axis), so that each point of the curve maps percentiles of the two distributions.

[Figure 2 here]

Taking the 5th percentile of the predicted approval probability of actually approved applicants as a threshold to classify discouraged borrowers as approved or denied, we find that about 21% of discouraged firms fall below this threshold and would likely be rejected. Only a quite small fraction of the discouraged firms in the sample thus correctly anticipate rejection. Even if we extend the threshold to the first quartile of approved applicants, we still observe that about 56% of discouraged firms have a higher approval probability and hence would likely have obtained credit if they had applied. This evidence suggests that the observed self-rationing mechanism is rather inefficient and hints at the potential role that financial knowledge might play in reducing demand-side constraints to micro-enterprises' access to credit.

**4. Conclusions**

This paper examines the role of financial knowledge on borrower discouragement among Italian micro-enterprises. We provide first empirical evidence that financial knowledge significantly reduces the probability of being discouraged due to complex application procedures and expected rejection. Our findings are robust to several sensitivity checks, including accounting for potential endogeneity, and allow us to shed light on the inefficiency of the observed self-rationing mechanism.

Taken together, the results obtained suggest that policy interventions aimed at enhancing financial knowledge among entrepreneurs might be effective to alleviate self-imposed financing constraints and reduce credit market imperfections, enhancing MSMEs' financial inclusion.

---

[5] Table S5 in the Supplementary Appendix presents results of the loan approval model. The ROC curve analysis (Figure S1) provides support to model accuracy: the area under the curve (AUROC) is significantly different from 0.5 and indicates that the model correctly predicts loan approval 76.5% of the time.

# References

Balana, B. B., Mekonnen, D., Haile, B., Hagos, F., Yimam, S. M., and Ringler, C. (2022). Demand and supply constraints of credit in smallholder farming: evidence from Ethiopia and Tanzania. World Dev. 159:106033.

Basha, S.A., Bennasr, H., Mohamed Goaied, M., 2023. Financial literacy, financial development, and leverage of small firms. Int. Rev. Fin. Anal. 86, 102510.

Bertrand, J., Burietz, A., Perrin, C., 2022. Just the two of us, we can('t) make it if we try: Owner-CEO gender and discouragement. Econ. Lett. 216, 110596.

Cole, R., Sokolyk, T., 2016. Who needs credit and who gets credit? Evidence from the surveys of small business finances. J. Financ. Stab. 24, 40–60.

Marc Cowling, M., Sclip, A., 2023. Dynamic Discouraged Borrowers. Br. J. Manag. 34, 1774–1790.

D'Ignazio, A., Finaldi Russo, P., Stacchini, M., 2022. Micro-entrepreneurs' financial and digital com- petences during the pandemic in Italy. Bank of Italy Occasional Paper 724, available at: https://doi.org/10.2139/ssrn.4463015.

Ferrando, A., Mulier, K., 2022. The real effects of credit constraints: Evidence from discouraged borrowers. J. Corp. Fin. 73, 102171.

Ferrando, A., Popov, A., Udell, G.F., 2017. Sovereign stress and SMEs' access to finance: Evidence from ECB's SAFE survey. J. Bank. Finance 81, 65–80.

Jappelli, T., 1990. Who is credit constrained in the US economy? Quart. J. Econ. 105, 219–234.

Kon, Y., Storey, D., 2003. A theory of discouraged borrowers. Small Business Econ. 21, 37–49.

Levenson, A.R. Willard, K.L., 2000. Do Firms Get the Financing They Want? Measuring Credit Rationing Experienced by Small Businesses in the US. Small Business Econ. 14(2), 83–94.

Lusardi, A., Mitchell, O. S., 2014. The economic importance of financial literacy: Theory and evidence. American Economic Journal: J. Econ. Lit. 52(1), 5–44.

Naegels, V., Mori, N., D'Espallier, B., 2022. The process of female borrower discouragement. Emerg. Mark. Rev. 50, 100837.

Nguyen, H.T., Nguyen, H.M., Troege, M., Nguyen, A.T.H., 2021. Debt aversion, education, and credit self-rationing in SMEs. Small Business Econ. 57, 1125–1143.

OECD, 2020. OECD/INFE survey instrument to measure the financial literacy of MSMEs, available at: https://www.oecd.org/financial/education/2020-survey-to-measure-msme-financial-literacy.pdf.

Stiglitz, J., Weiss, A., 1981. Credit rationing in markets with imperfect information. Am. Econ. Rev. 71, 393–410.

Wernli, R., Dietrich, A., 2022. Only the brave: improving self-rationing efficiency among discouraged Swiss SMEs. Small Business Econ. 59, 977–1003.

Xu, N., Shi, J., Rong, Z., Yuan, Y., 2020. Financial literacy and formal credit accessibility: Evidence from informal businesses in China. Fin. Res. Lett. 36, 101327.

# Tables

Table 1 – Variable definitions

| Variable | Definition | Mean |
|---|---|---|
| *a) Dependent variables* | | |
| Need | Equals 1 if the firm, since the start of the COVID-19 pandemic, either applied for a loan or did not apply despite it needed new financing; 0 otherwise | 0.6816 |
| Discouraged | Equals 1 if the firm, since the start of the COVID-19 pandemic, needed new financing, but did not apply for a bank loan due to too complex application procedures or because it did not think the loan would be approved; 0 otherwise | 0.1261 |
| *a) Explanatory variables* | | |
| Financial knowledge (FK) | Number of correct answers to seven test-based questions related to: 1) simple interest; 2) compound interest; 3) risk-return relationship; 4) inflation; 5) interest on loans; 6) credit ratings; and 7) risk diversification | 5.0114 |
| Age of the firm | Age of the firm (in logs) | 2.1897 |
| Autonomous firm | Equals 1 if the firm is an autonomous profit-oriented business, making independent financial decisions; 0 otherwise | 0.9496 |
| One-person firm | Equals 1 if one full-time equivalent person, including the owner, works in the firm; 0 otherwise | 0.0919 |
| Turnover: €100,000-€500,000 | Equals 1 if the firm's turnover is more than €100,000 and up to €500,000; 0 otherwise | 0.5010 |
| Turnover: More than €500,000 | Equals 1 if the firm's turnover is more than €500,000; 0 otherwise | 0.2909 |
| Exporter | Equals 1 if the firm exports products or offers services abroad; 0 otherwise | 0.1560 |
| Female owner | Equals 1 if the firm's owner is a woman; 0 otherwise | 0.2845 |
| Age of the owner | Age of the firm's owner (in logs) | 3.8792 |
| More than 10 years of experience | Equals 1 if the firm's owner has more than 10 years of entrepreneurial experience; 0 otherwise | 0.5715 |
| Upper secondary education | Equals 1 if the firm's owner has an upper secondary education; 0 otherwise | 0.5645 |
| University-level education | Equals 1 if the firm's owner has a university-level education; 0 otherwise | 0.2322 |
| Post-graduate education | Equals 1 if the firm's owner has a post-graduate education; 0 otherwise | 0.0586 |
| Received help accessing finance | Equals 1 if the firm's owner, during the last 24 months, asked for help about accessing external finance; 0 otherwise | 0.4565 |
| Low pre-pandemic liquidity | Equals 1 if the firm, at the end of 2019, had a quite low or too low liquidity; 0 otherwise | 0.3445 |
| Liquidity decreased | Equals 1 if the firm experienced a decrease in liquidity due to the impact of the COVID-19 crisis; 0 otherwise | 0.5337 |
| Profits decreased | Equals 1 if the firm experienced a decrease in profits due to the impact of the COVID-19 crisis; 0 otherwise | 0.6636 |
| Centre | Equals 1 if the firm is located in Central Italy; 0 otherwise | 0.2117 |
| South & Islands | Equals 1 if the firm is located in Southern Italy or Islands; 0 otherwise | 0.2902 |

**Notes**: the average values of all the dependent and explanatory variables, computed using sample weights.

Table 2 – Baseline results

| Dependent variable: | Discouraged (1) | Need (2) |
|---|---|---|
| Financial knowledge | -0.0148*** | -0.0078 |
|  | (0.0036) | (0.0066) |
| Age of the firm (in logs) | -0.0105 | 0.0040 |
|  | (0.0171) | (0.0072) |
| Autonomous firm | 0.0115 | 0.0176 |
|  | (0.0281) | (0.0492) |
| One-person firm | 0.0756** | -0.0408** |
|  | (0.0348) | (0.0170) |
| Turnover: €100,000-€500,000 | -0.0403* | -0.0005 |
|  | (0.0239) | (0.0314) |
| Turnover: More than €500,000-€1 million | -0.0330*** | 0.0003 |
|  | (0.0100) | (0.0332) |
| Exporter | -0.0144 | 0.0403* |
|  | (0.0214) | (0.0225) |
| Female owner | -0.0380** | 0.0042 |
|  | (0.0186) | (0.0215) |
| Age of the owner (in logs) | 0.0325 | -0.0264 |
|  | (0.0415) | (0.0476) |
| More than 10 years of experience | -0.0125 | 0.0219 |
|  | (0.0201) | (0.0216) |
| Upper secondary education | -0.0603** | -0.0343 |
|  | (0.0282) | (0.0325) |
| University-level education | -0.0406 | -0.1012*** |
|  | (0.0317) | (0.0369) |
| Post-graduate education | -0.0026 | -0.0787* |
|  | (0.0470) | (0.0420) |
| Centre | 0.0370** | 0.0833*** |
|  | (0.0184) | (0.0221) |
| South & Islands | 0.0756*** | 0.0731*** |
|  | (0.0164) | (0.0220) |
| Received help accessing finance | -0.0995*** | 0.2519*** |
|  | (0.0222) | (0.0254) |
| Low pre-pandemic liquidity |  | 0.1999*** |
|  |  | (0.0212) |
| Liquidity decreased |  | 0.1159*** |
|  |  | (0.0323) |
| Profits decreased |  | 0.0465* |
|  |  | (0.0263) |
| Sector fixed effects | Yes | Yes |
| Error correlation ($\rho$) | -0.6350*** |  |
|  | (0.0651) |  |
| Log-likelihood | -1320.94 |  |
| Observations | 1,821 |  |

Notes: the Table reports estimated average marginal effects on the conditional probability that a firm did not apply for a loan (due to either too complex application procedures or because it expected rejection) and on the probability of needing a loan, estimated from a probit model with endogenous sample selection. Standard errors, clustered at the sector level, are reported in parentheses below the estimates.

***, ** and * denote significance at the 1, 5 and 10% levels, respectively

Table 3 – Robustness checks and additional analyses

*Panel a) Controlling for the potential endogeneity of FK*

| Dependent variable: | Discouraged (1) | Need (2) |
|---|---|---|
| Financial knowledge | -0.0193*** | -0.0093 |
|  | (0.0050) | (0.0217) |
| Error correlation ($\rho$) | -0.6302*** |  |
|  | (0.0668) |  |
| Wald test of exogeneity | 0.04 | 0.00 |
|  | [0.8484] | [0.9518] |
| Overidentification test | 0.43 |  |
|  | [0.4864] |  |
| Weak-instrument F test | 104.18 |  |
|  | [0.0000] |  |
| Log-likelihood | -4854.11 |  |
| Observations | 1,821 |  |

*Panel b) Disentangling the reasons of borrower discouragement*

| Dependent variable: | Too complex procedures (1) | Need (2) | Expect rejection (3) | Need (4) |
|---|---|---|---|---|
| Financial knowledge | -0.0075*** | -0.0056 | -0.0106*** | -0.0082 |
|  | (0.0022) | (0.0062) | (0.0038) | (0.0067) |
| Error correlation ($\rho$) | -0.4822*** |  | -0.6595*** |  |
|  | (0.1350) |  | (0.1303) |  |
| Log-likelihood | -1171.32 |  | -1092.34 |  |
| Observations | 1,755 |  | 1,730 |  |

*Panel c) Controlling for business finance and accounting knowledge*

| Dependent variable: | Credit discouraged (1) | Need (2) | Too complex procedures (3) | Need (4) | Expect rejection (5) | Need (6) |
|---|---|---|---|---|---|---|
| Financial knowledge | -0.0176*** | -0.0005 | -0.0107*** | 0.0018 | -0.0114** | 0.0001 |
|  | (0.0053) | (0.0071) | (0.0040) | (0.0070) | (0.0050) | (0.0070) |
| Bus. finance/accounting knowledge | 0.0080 | -0.0231*** | 0.0091 | -0.0229*** | 0.0021 | -0.0264*** |
|  | (0.0111) | (0.0084) | (0.0113) | (0.0089) | (0.0070) | (0.0090) |
| Error correlation ($\rho$) | -0.6374*** |  | -0.4823*** |  | -0.6607*** |  |
|  | (0.0633) |  | (0.1362) |  | (0.1281) |  |
| Log-likelihood | -1317.94 |  | -1168.34 |  | -1089.13 |  |
| Observations | 1,821 |  | 1,755 |  | 1,730 |  |

Notes: the Table reports estimated average marginal effects on the conditional probability that a firm did not apply for a loan (due to any reason, to too complex application procedures, and because it expected rejection) and on the probability of needing a loan, estimated from probit models with endogenous sample selection. All the models include the same control variables included in the baseline specification (see Table 2). In panel a), we use as additional instrumental variables: 1) a dummy indicating whether the entrepreneur received education in subjects related to business, economics, or finance as part of your school or university education; 2) a dummy indicating whether at least one of the entrepreneur's parents currently own or owned a business in the past. The p-values of the Wald tests of exogeneity, of the overidentification test, and of the F test for weak instruments are reported in square brackets. Standard errors, clustered at the sector level, are reported in parentheses below the estimates.

***, ** and * denote significance at the 1, 5 and 10% levels, respectively

**Figures**

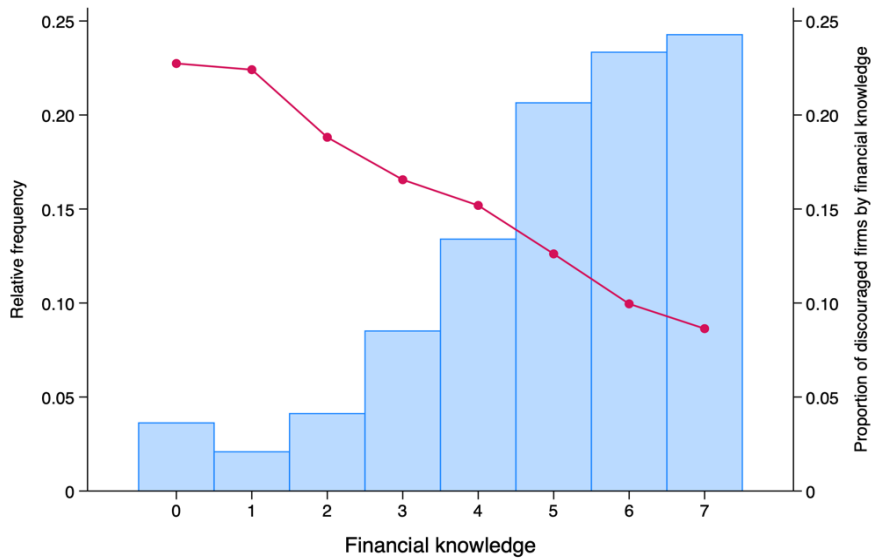

Figure 1 – Financial knowledge and borrower discouragement

**Note**: The histogram describes the observed distribution of entrepreneurs' financial knowledge (left scale), while the line plot reports the proportion of discouraged borrowers by level financial knowledge (right scale).

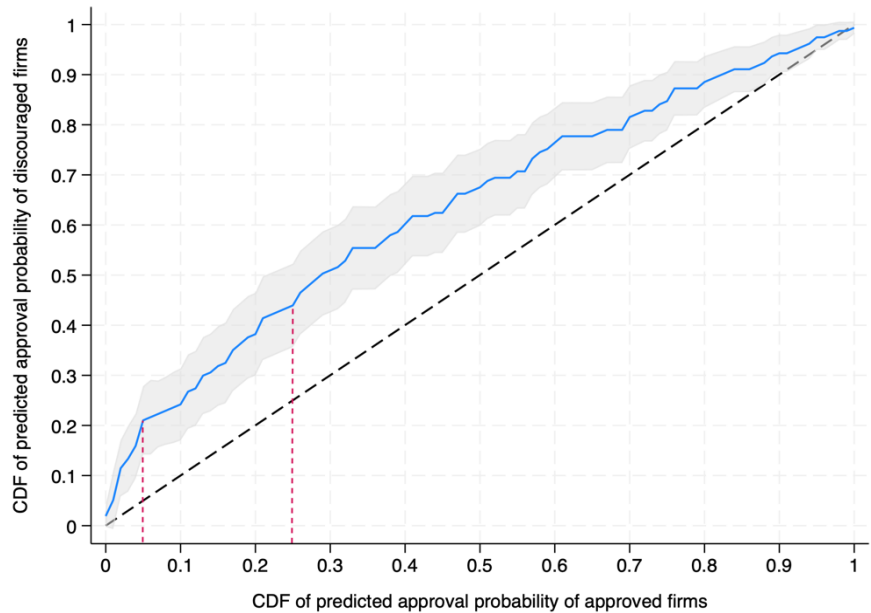

Figure 2 – Relative distribution of predicted approval probability of discouraged firms

**Note**: The Figure presents the cumulative distribution function of the predicted approval probability of discouraged firms relative to that of approved firms (based on results of the loan approval model presented in Table S5 of the Supplementary Appendix), with bootstrapped (500 reps) 95% confidence intervals. The distribution of the predicted approval probability of discouraged firms is equivalent to that of approved firms when the relative CDF lies on the 45 degrees line.

# Financial knowledge and borrower discouragement

**Supplementary Appendix**

Table S1 – Questions on financial knowledge and business finance/accounting knowledge

| | |
|---|---|
| a) Financial knowledge | |
| QK3 | *Imagine that someone puts €100 into a <no fee, tax free> savings account with a guaranteed interest rate of 2% per year. They don't make any further payments into this account and they don't withdraw any money. How much would be in the account at the end of the first year, once the interest payment is made?*<br>**102;** -97=Don't know; -99=Refused. |
| QK4 | *…and how much would be in the account at the end of five years [add if necessary: remembering there are no fees or tax deductions]? Would it be:*<br>**1=More than €110;** 2=Exactly €110; 3=Less than €110; 4=Impossible to tell from the information given; -97=Don't know; -99=Refused. |
| QK7_3 | *If a financial investment offers the chance to make a lot of money it is likely that there is also a chance to lose a lot of money.*<br>**1=True;** 0=False; -97=Don't know; -99=Refused. |
| QK7_4 | *High inflation means that the cost of living is increasing rapidly.*<br>**1=True;** 0=False; -97=Don't know; -99=Refused. |
| QK7_5 | *A 15-year loan typically requires higher monthly payments than a 30-year loan, but the total interest paid over the life of the loan will be less.*<br>**1=True;** 0=False; -97=Don't know; -99=Refused. |
| QK7_6 | *Credit rating is an evaluation of the ability of a prospective borrower to pay back their debt*<br>**1=True;** 0=False; -97=Don't know; -99=Refused. |
| QK7_7 | ***If a farmer grows several types of fruit and vegetables each year, she has a lower risk of losing all her crops to disease***<br>**1=True;** 0=False; -97=Don't know; -99=Refused. |
| b) Business finance and accounting knowledge | |
| QK5 | *Could you tell me which of these best describes a balance sheet?*<br>1=**A financial snapshot, taken at a point in time, of the firm's assets and liabilities;** 2=A record of profits and losses of the firm in a certain period of time; 3=A record of the flow of financial resources over time; 4=None of those; -97=Don't know; -99=Refused. |
| QK6 | *Could you tell me which of these best describes the Return-on-Assets ratio (ROA)?*<br>1=An indicator of the firm's capital structure; 2=An indicator of the firm's liquidity; **3=An indicator of the firm's performance;** 4=None of those; -97=Don't know; -99=Refused. |
| QK7_1 | *Dividends are part of what a business pays to a bank to repay a loan.*<br>1=True; **0=False;** -97=Don't know; -99=Refused. |
| QK7_2 | *When a company obtains equity from an investor it gives the investor part of the ownership of the company.*<br>**1=True;** 0=False; -97=Don't know; -99=Refused. |

**Notes**: Questions are drawn from the questionnaire developed by OECD-INFE (OECD, 2020). Correct answers are in bold.

Table S2 – Controlling for the potential endogeneity of financial knowledge

| Dependent variable: | Discouraged (1) | Need (2) | Financial knowledge (3) |
|---|---|---|---|
| Financial knowledge | -0.0193*** | -0.0093 | |
| | (0.0050) | (0.0217) | |
| Age of the firm (in logs) | -0.0108 | 0.0039 | -0.0729* |
| | (0.0173) | (0.0071) | (0.0379) |
| Autonomous firm | 0.0163 | 0.0193 | 1.1031*** |
| | (0.0252) | (0.0551) | (0.1676) |
| One-person firm | 0.0740** | -0.0412** | -0.1771 |
| | (0.0338) | (0.0165) | (0.1679) |
| Turnover: €100,000-€500,000 | -0.0389* | -0.0000 | 0.3028** |
| | (0.0236) | (0.0352) | (0.1176) |
| Turnover: More than €500,000 | -0.0318*** | 0.0007 | 0.3206** |
| | (0.0094) | (0.0375) | (0.1470) |
| Exporter | -0.0138 | 0.0405* | 0.1455 |
| | (0.0211) | (0.0230) | (0.1008) |
| Female owner | -0.0390** | 0.0039 | -0.2579** |
| | (0.0182) | (0.0251) | (0.1030) |
| Age of the owner (in logs) | 0.0369 | -0.0248 | 1.3686*** |
| | (0.0433) | (0.0447) | (0.2152) |
| More than 10 years of experience | -0.0127 | 0.0219 | -0.1391 |
| | (0.0198) | (0.0214) | (0.1016) |
| Upper secondary education | -0.0578** | -0.0335 | 0.3053* |
| | (0.0290) | (0.0368) | (0.1624) |
| University-level education | -0.0371 | -0.0998** | 0.5444*** |
| | (0.0319) | (0.0408) | (0.1498) |
| Post-graduate education | 0.0020 | -0.0772* | 0.7136*** |
| | (0.0475) | (0.0420) | (0.2045) |
| Centre | 0.0362* | 0.0830*** | -0.1467 |
| | (0.0185) | (0.0221) | (0.1155) |
| South & Islands | 0.0750*** | 0.0728*** | -0.1129 |
| | (0.0164) | (0.0208) | (0.1219) |
| Received help accessing finance | -0.0984*** | 0.2523*** | 0.3134*** |
| | (0.0229) | (0.0248) | (0.0741) |
| Low pre-pandemic liquidity | | 0.2001*** | 0.0389 |
| | | (0.0215) | (0.0850) |
| Liquidity decreased | | 0.1153*** | -0.1723*** |
| | | (0.0300) | (0.0415) |
| Profits decreased | | 0.0463* | -0.0974 |
| | | (0.0280) | (0.0865) |
| Financial education at school | | | 0.4441*** |
| | | | (0.0606) |
| Parents' entrepreneurial experience | | | 0.5079*** |
| | | | (0.0786) |
| Sector fixed effects | Yes | Yes | Yes |
| Error correlation (ρ) | -0.6302*** | | |
| | (0.0668) | | |
| Wald test of exogeneity | 0.04 | 0.00 | |
| | [0.8484] | [0.9518] | |
| Overidentification test | 0.43 | | |
| | [0.4864] | | |
| Weak-instrument F test | 104.18 | | |
| | [0.0000] | | |
| Log-likelihood | -4854.11 | | |
| Observations | 1,821 | | |

Notes: the Table reports estimated average marginal effects on the conditional probability that a firm did not apply for a loan (due to either too complex application procedures or because it expected rejection), on the probability of needing a loan, and on the number of correct answers to financial knowledge questions, estimated from a probit model with endogenous sample selection, extended to account for the potential endogeneity of FK. Standard errors, clustered at the sector level, are reported in parentheses below the estimates. The p-values of the Wald tests of exogeneity, of the overidentification test, and of the F test for weak instruments are reported in square brackets.

***, ** and * denote significance at the 1, 5 and 10% levels, respectively

Table S3 – Disentangling the reasons of borrower discouragement

| Dependent variable: | Too complex procedures (1) | Need (2) | Expect rejection (3) | Need (4) |
|---|---|---|---|---|
| Financial knowledge | -0.0075*** | -0.0056 | -0.0106*** | -0.0082 |
|  | (0.0022) | (0.0062) | (0.0038) | (0.0067) |
| Age of the firm (in logs) | 0.0069 | 0.0044 | -0.0167* | 0.0031 |
|  | (0.0121) | (0.0081) | (0.0098) | (0.0083) |
| Autonomous firm | 0.0089 | 0.0232 | 0.0108 | 0.0359 |
|  | (0.0388) | (0.0528) | (0.0170) | (0.0477) |
| One-person firm | 0.0572** | -0.0466** | 0.0336 | -0.0510*** |
|  | (0.0287) | (0.0184) | (0.0214) | (0.0154) |
| Turnover: €100,000-€500,000 | -0.0252 | 0.0032 | -0.0229 | 0.0030 |
|  | (0.0229) | (0.0311) | (0.0153) | (0.0266) |
| Turnover: More than €500,000 | -0.0270 | -0.0009 | -0.0085 | 0.0068 |
|  | (0.0210) | (0.0347) | (0.0136) | (0.0328) |
| Exporter | -0.0276 | 0.0437* | 0.0089 | 0.0421* |
|  | (0.0173) | (0.0243) | (0.0318) | (0.0223) |
| Female owner | -0.0457*** | 0.0071 | 0.0008 | 0.0125 |
|  | (0.0135) | (0.0251) | (0.0186) | (0.0224) |
| Age of the owner (in logs) | 0.0192 | -0.0351 | 0.0355 | -0.0246 |
|  | (0.0350) | (0.0540) | (0.0223) | (0.0508) |
| More than 10 years of experience | -0.0030 | 0.0254 | -0.0142 | 0.0195 |
|  | (0.0188) | (0.0213) | (0.0128) | (0.0185) |
| Upper secondary education | -0.0235 | -0.0361 | -0.0426** | -0.0281 |
|  | (0.0251) | (0.0321) | (0.0166) | (0.0334) |
| University-level education | -0.0264 | -0.1076*** | -0.0197 | -0.0915*** |
|  | (0.0286) | (0.0389) | (0.0161) | (0.0341) |
| Post-graduate education | 0.0102 | -0.0785* | -0.0121 | -0.0803** |
|  | (0.0423) | (0.0407) | (0.0232) | (0.0378) |
| Centre | 0.0018 | 0.0817*** | 0.0320** | 0.0802*** |
|  | (0.0176) | (0.0235) | (0.0162) | (0.0206) |
| South & Islands | 0.0581*** | 0.0672*** | 0.0285*** | 0.0606** |
|  | (0.0196) | (0.0218) | (0.0097) | (0.0254) |
| Received help accessing finance | -0.0815*** | 0.2645*** | -0.0383** | 0.2773*** |
|  | (0.0305) | (0.0252) | (0.0164) | (0.0281) |
| Low pre-pandemic liquidity |  | 0.1902*** |  | 0.2146*** |
|  |  | (0.0217) |  | (0.0241) |
| Liquidity decreased |  | 0.1174*** |  | 0.1053*** |
|  |  | (0.0350) |  | (0.0390) |
| Profits decreased |  | 0.0551** |  | 0.0282 |
|  |  | (0.0266) |  | (0.0293) |
| Sector fixed effects | Yes | Yes | Yes | Yes |
| Error correlation ($\rho$) | -0.4822*** |  | -0.6595*** |  |
|  | (0.1350) |  | (0.1303) |  |
| Log-likelihood | -1171.32 |  | -1092.34 |  |
| Observations | 1,755 |  | 1,730 |  |

Notes: the Table reports estimated average marginal effects on the conditional probability that a firm did not apply for a loan due to too complex application procedures, on the conditional probability that a firm did not apply because it expected rejection, and on the probability of needing a loan, estimated from probit models with endogenous sample selection. Standard errors, clustered at the sector level, are reported in parentheses below the estimates.
***, ** and * denote significance at the 1, 5 and 10% levels, respectively

Table S4 – Controlling for business and accounting knowledge

| Dependent variable: | Discouraged (1) | Need (2) | Too complex procedures (3) | Need (4) | Expect rejection (5) | Need (6) |
|---|---|---|---|---|---|---|
| Financial knowledge | -0.0176*** | -0.0005 | -0.0107*** | 0.0018 | -0.0114** | 0.0001 |
| | (0.0053) | (0.0071) | (0.0040) | (0.0070) | (0.0050) | (0.0070) |
| Bus. finance/accounting knowledge | 0.0080 | -0.0231*** | 0.0091 | -0.0229** | 0.0021 | -0.0264*** |
| | (0.0110) | (0.0084) | (0.0113) | (0.0089) | (0.0070) | (0.0090) |
| Age of the firm (in logs) | -0.0102 | 0.0033 | 0.0072 | 0.0038 | -0.0167* | 0.0023 |
| | (0.0169) | (0.0072) | (0.0119) | (0.0079) | (0.0098) | (0.0083) |
| Autonomous firm | 0.0073 | 0.0262 | 0.0045 | 0.0317 | 0.0097 | 0.0445 |
| | (0.0304) | (0.0488) | (0.0416) | (0.0524) | (0.0164) | (0.0481) |
| One-person firm | 0.0749** | -0.0414** | 0.0573* | -0.0479** | 0.0333 | -0.0524*** |
| | (0.0341) | (0.0174) | (0.0295) | (0.0191) | (0.0207) | (0.0163) |
| Turnover: €100,000-€500,000 | -0.0412* | 0.0020 | -0.0263 | 0.0058 | -0.0232 | 0.0061 |
| | (0.0239) | (0.0318) | (0.0224) | (0.0315) | (0.0154) | (0.0274) |
| Turnover: More than €500,000 | -0.0342*** | 0.0036 | -0.0289 | 0.0027 | -0.0086 | 0.0101 |
| | (0.0104) | (0.0336) | (0.0205) | (0.0353) | (0.0133) | (0.0336) |
| Exporter | -0.0145 | 0.0394* | -0.0273 | 0.0429* | 0.0086 | 0.0413* |
| | (0.0211) | (0.0224) | (0.0176) | (0.0239) | (0.0319) | (0.0223) |
| Female owner | -0.0392** | 0.0077 | -0.0469*** | 0.0106 | 0.0004 | 0.0166 |
| | (0.0172) | (0.0209) | (0.0134) | (0.0242) | (0.0179) | (0.0218) |
| Age of the owner (in logs) | 0.0323 | -0.0209 | 0.0195 | -0.0300 | 0.0355 | -0.0185 |
| | (0.0418) | (0.0479) | (0.0356) | (0.0542) | (0.0224) | (0.0520) |
| More than 10 years of experience | -0.0137 | 0.0256 | -0.0045 | 0.0292 | -0.0145 | 0.0237 |
| | (0.0191) | (0.0226) | (0.0187) | (0.0224) | (0.0130) | (0.0196) |
| Upper secondary education | -0.0624** | -0.0307 | -0.0245 | -0.0334 | -0.0437*** | -0.0235 |
| | (0.0277) | (0.0320) | (0.0252) | (0.0317) | (0.0156) | (0.0329) |
| University-level education | -0.0420 | -0.0957*** | -0.0273 | -0.1026*** | -0.0204 | -0.0846** |
| | (0.0311) | (0.0371) | (0.0289) | (0.0392) | (0.0148) | (0.0342) |
| Post-graduate education | -0.0057 | -0.0703* | 0.0082 | -0.0716* | -0.0133 | -0.0699** |
| | (0.0476) | (0.0393) | (0.0433) | (0.0383) | (0.0221) | (0.0346) |
| Centre | 0.0365** | 0.0869*** | 0.0006 | 0.0850*** | 0.0322** | 0.0837*** |
| | (0.0182) | (0.0207) | (0.0178) | (0.0226) | (0.0161) | (0.0192) |
| South & Islands | 0.0758*** | 0.0735*** | 0.0581*** | 0.0675*** | 0.0288*** | 0.0603** |
| | (0.0164) | (0.0211) | (0.0196) | (0.0211) | (0.0102) | (0.0238) |
| Received help accessing finance | -0.1002*** | 0.2534*** | -0.0818*** | 0.2655*** | -0.0387** | 0.2795*** |
| | (0.0218) | (0.0253) | (0.0296) | (0.0250) | (0.0157) | (0.0281) |
| Low pre-pandemic liquidity | | 0.1991*** | | 0.1896*** | | 0.2137*** |
| | | (0.0208) | | (0.0218) | | (0.0236) |
| Liquidity decreased | | 0.1147*** | | 0.1164*** | | 0.1039*** |
| | | (0.0324) | | (0.0352) | | (0.0395) |
| Profits decreased | | 0.0473* | | 0.0559** | | 0.0287 |
| | | (0.0258) | | (0.0259) | | (0.0287) |
| Sector fixed effects | Yes | Yes | Yes | Yes | Yes | Yes |
| Error correlation ($\rho$) | -0.6374*** | | -0.4823*** | | -0.6607*** | |
| | (0.0633) | | (0.1362) | | (0.1281) | |
| Log-likelihood | -1317.94 | | -1168.34 | | -1089.13 | |
| Observations | 1,821 | | 1,755 | | 1,730 | |

Notes: the Table reports estimated average marginal effects on the conditional probability that a firm did not apply for a loan (due to any reason, to too complex application procedures, and because it expected rejection) and on the probability of needing a loan, estimated from a probit model with endogenous sample selection. Standard errors, clustered at the sector level, are reported in parentheses below the estimates.

***, ** and * denote significance at the 1, 5 and 10% levels, respectively

Table S5 – The determinants of loan approval

| Dependent variable: | Granted (1) | Apply (2) | Need (3) |
|---|---|---|---|
| Financial knowledge | -0.0068 | 0.0147*** | -0.0077 |
|  | (0.0046) | (0.0036) | (0.0066) |
| Age of the firm (in logs) | 0.0392*** | 0.0108 | 0.0061 |
|  | (0.0079) | (0.0168) | (0.0071) |
| Autonomous firm | 0.0252 | -0.0122 | 0.0213 |
|  | (0.0379) | (0.0286) | (0.0495) |
| One-person firm | -0.0259 | -0.0763** | -0.0403** |
|  | (0.0350) | (0.0346) | (0.0180) |
| Turnover: €100,000-€500,000 | 0.0263* | 0.0397* | 0.0042 |
|  | (0.0148) | (0.0238) | (0.0306) |
| Turnover: More than €500,000 | -0.0029 | 0.0318*** | 0.0047 |
|  | (0.0241) | (0.0097) | (0.0306) |
| Exporter | -0.0302* | 0.0132 | 0.0394* |
|  | (0.0183) | (0.0214) | (0.0206) |
| Female owner | -0.0283* | 0.0390** | 0.0048 |
|  | (0.0170) | (0.0186) | (0.0201) |
| Age of the owner (in logs) | -0.0305 | -0.0326 | -0.0284 |
|  | (0.0652) | (0.0407) | (0.0496) |
| More than 10 years of experience | -0.0189 | 0.0117 | 0.0238 |
|  | (0.0144) | (0.0203) | (0.0203) |
| Upper secondary education | 0.0146 | 0.0586** | -0.0385 |
|  | (0.0210) | (0.0288) | (0.0318) |
| University-level education | 0.0140 | 0.0407 | -0.1053*** |
|  | (0.0190) | (0.0324) | (0.0378) |
| Post-graduate education | -0.0356** | 0.0026 | -0.0878* |
|  | (0.0152) | (0.0504) | (0.0455) |
| Centre | -0.0156 | -0.0357** | 0.0836*** |
|  | (0.0098) | (0.0179) | (0.0227) |
| South & Islands | -0.0185 | -0.0760*** | 0.0744*** |
|  | (0.0160) | (0.0162) | (0.0203) |
| High pre-pandemic interest expenses | -0.0482** |  |  |
|  | (0.0203) |  |  |
| High pre-pandemic financial debts | -0.0315** |  |  |
|  | (0.0131) |  |  |
| Public support | 0.0881*** |  |  |
|  | (0.0171) |  |  |
| Received help accessing finance |  | 0.1004*** | 0.2509*** |
|  |  | (0.0225) | (0.0262) |
| Low pre-pandemic liquidity |  |  | 0.1943*** |
|  |  |  | (0.0209) |
| Liquidity decreased |  |  | 0.1283*** |
|  |  |  | (0.0316) |
| Profits decreased |  |  | 0.0415 |
|  |  |  | (0.0278) |
| Sector fixed effects | Yes | Yes | Yes |
| Error correlation coefficients: |  |  |  |
| $\rho_{GA}$ | -0.0778 |  |  |
|  | (0.2687) |  |  |
| $\rho_{GN}$ | 0.6302*** |  |  |
|  | (0.1298) |  |  |
| $\rho_{AN}$ | 0.6309*** |  |  |
|  | (0.0681) |  |  |
| Log-likelihood | -1619.00 |  |  |
| Observations | 1,821 |  |  |

Notes: the Table reports average marginal effects on the conditional probability of having fully obtained a loan, on the conditional probability of having applied for a loan, and on the probability of needing a loan, estimated from a probit model with double sample selection. *High pre-pandemic interest expenses* and *High pre-pandemic financial debts* are equal to 1 if the firm, at the end of 2019, had high interest expenses and financial debts, respectively; *Public support* equals 1 if the firm received public financial support. Standard errors, clustered at the sector level, are reported in parentheses below the estimates.
\*\*\*, \*\* and \* denote significance at the 1, 5 and 10% levels, respectively

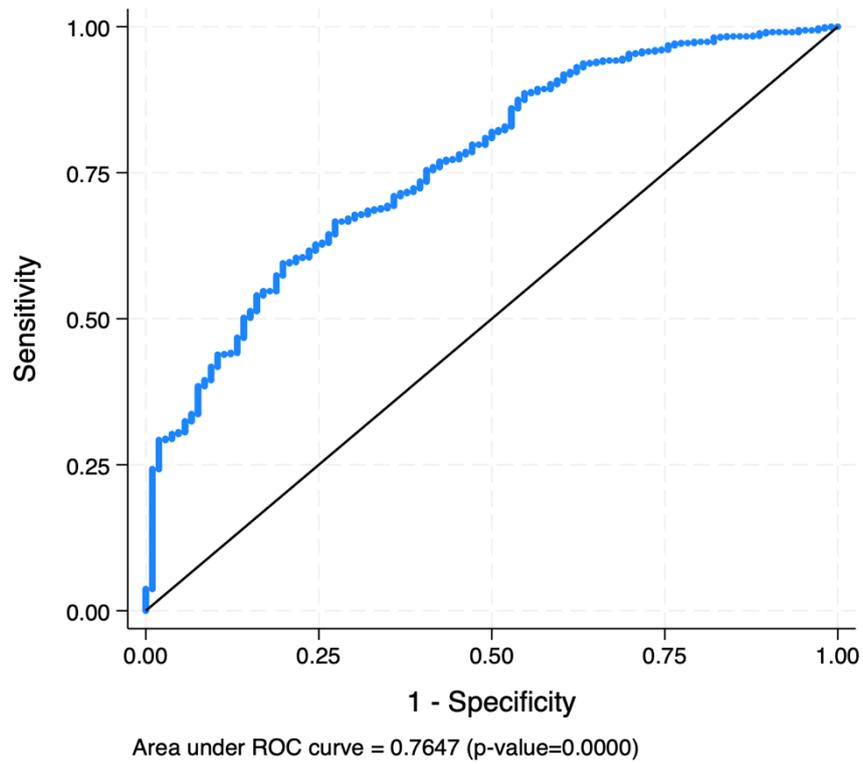

Figure S1 – Prediction accuracy of the loan approval model

**Note**: The Figure shows the Receiver Operating Characteristics (ROC) curve, together with the corresponding area under the curve (AUROC), based on results of the loan approval model presented in Table S5.